\documentclass[aps,prl,reprint,showpacs,byrevtex]{revtex4-2}
\let\oldhat\hat
\renewcommand{\hat}[1]{\oldhat{\mathbf{#1}}}
\usepackage{titlesec}
\usepackage{array}

\usepackage{times,mathptm}% for making `pdf' file
\usepackage[dvips]{graphicx,color}% for inclusion of graphics file(s)

\usepackage{amsmath}
\usepackage{xcolor}

\begin{document}
\title{Contrasting magnetic anisotropy in CrCl$_3$ and CrBr$_3$: A first-principles study}
\author{Jiazhuang Si$^{1}$, Shuyuan Liu$^{2}$, Bing Wang$^{1*}$, Chongze Wang$^2$, Fengzhu Ren$^1$, Yu Jia$^{3,4}$, and Jun-Hyung Cho$^{1,2*}$}
\affiliation{$^1$Joint Center for Theoretical Physics, School of Physics and Electronics, Henan University, Kaifeng 475004, China \\
$^2$Department of Physics and Research Institute for Natural Science, Hanyang University, 222 Wangsimni-ro, Seongdong-Ku, Seoul 04763, Repu\emph{}blic of Korea \\
$^3$Key Laboratory for Special Functional Materials of the Ministry of Education, Henan University, Kaifeng 475004, China \\
$^4$Institute of Quantum Materials and Physics, Henan Academy of Sciences, Zhengzhou 450046, China}
\date{\today}

\begin{abstract}
We present a first-principles study of the contrasting easy magnetization axes (EMAs) in the layered chromium trihalides CrCl$_3$ and CrBr$_3$, which exhibit in-plane and out-of-plane EMAs, respectively. Using density-functional theory calculations, we show that the EMA is determined by the interplay between spin–orbit coupling-induced magnetocrystalline anisotropy energy (SOC-MAE) and shape magnetic anisotropy energy (shape-MAE) arising from dipole–dipole interactions. While the Cr $d$ orbitals contribute similarly to the SOC-MAE in both compounds, the key difference stems from the halogen $p$ orbitals. In CrCl$_3$, the localized Cl $3p$ orbitals and weak SOC favor spin-flip SOC interactions, particularly between the $(p_x, p_y)$ and $(p_y, p_z)$ channels. These channels contribute with opposite signs--negative and positive, respectively--leading to partial cancellation and a small net SOC-MAE. As a result, the shape-MAE exceeds the SOC-MAE in magnitude, favoring an in-plane EMA. In contrast, CrBr$_3$ features more delocalized Br $4p$ orbitals, stronger SOC, and enhanced $p$–$d$ hybridization. This leads to stronger spin-conserving SOC interactions, with dominant contributions from both the $(p_x, p_y)$ and $(p_y, p_z)$ channels. In this case, the positive contribution from the $(p_x, p_y)$ channel outweighs the smaller negative contribution from the $(p_y, p_z)$ channel, resulting in a sizable net SOC-MAE. The SOC-MAE thus surpasses the shape-MAE and stabilizes an out-of-plane EMA. These findings demonstrate that the contrasting magnetic anisotropies in CrCl$_3$ and CrBr$_3$ originate from differences in the spatial distribution, SOC strength, and hybridization of the halogen $p$ orbitals, highlighting the critical role of orbital anisotropy and spin selection rules in governing magnetic behavior in layered semiconductors.
\end{abstract}

\maketitle
%\begin{multicols}{2}

%\vspace{0.4cm}
\section{I. INTRODUCTION}
%\vspace{0.4cm}

Since the discovery of intrinsic ferromagnetism in atomically thin films of CrI$_3$~\cite{NatureCrI3} and Cr$_2$Ge$_2$Te$_6$~\cite{NatureCrGeTe3}, two-dimensional (2D) van der Waals (vdW) magnetic materials have emerged as a versatile platform for exploring magnetism in reduced dimensions and for enabling spin-based technologies at the nanoscale.~\cite{3,natureFeGeTe,2d vdw,vse2,NC FeGeTe,NM 2d,2d magnetic} These materials uniquely offer the ability to incorporate magnetic functionality into atomically thin layered structures, opening new possibilities for spintronic and magnetoelectronic applications~\cite{Nanotechnol-2019,science-2019}. Remarkably, many 2D vdW magnets preserve robust long-range magnetic order even down to the monolayer limit—an unexpected observation in light of the Mermin--Wagner theorem~\cite{PRL-MW1966}, which states that spontaneous symmetry breaking, and thus long-range magnetic order, is forbidden in perfectly isotropic two-dimensional systems at finite temperatures due to enhanced thermal fluctuations. This apparent contradiction is resolved by the presence of magnetic anisotropy energy (MAE), which energetically favors a specific magnetization direction, known as the easy magnetization axis (EMA), and consequently opens a spin-wave gap that suppresses low-energy thermal excitations. The resulting anisotropy stabilizes magnetic order against thermal agitation, thereby enabling finite-temperature ferromagnetism in 2D systems. Indeed, the relatively high Curie temperatures observed in CrI$_3$ and Cr$_2$Ge$_2$Te$_6$ monolayers have been attributed to their significant out-of-plane MAEs~\cite{NatureCrI3,NatureCrGeTe3}. A detailed understanding of the microscopic origin of MAE is therefore essential for both fundamental studies of 2D magnetism and the development of magnetic components in future quantum and spin-based electronic devices.

Among the family of 2D vdW magnets, layered chromium trihalides CrCl$_3$ and CrBr$_3$ stand out as prototypical systems for exploring magnetic anisotropy, owing to their structural simplicity, intrinsic ferromagnetism, and tunable magnetic properties via halogen substitution~\cite{AMCrCl3-xBrx}. Despite their close structural and electronic resemblance, these compounds exhibit strikingly different magnetic anisotropy: CrCl$_3$ prefers an in-plane EMA~\cite{NL-CrCl3}, whereas CrBr$_3$ favors an out-of-plane EMA~\cite{PNAS-CrX3}. Unraveling the microscopic origin of this contrasting behavior is crucial not only for advancing our fundamental understanding of halide-based layered magnets, but also for guiding the design of materials with engineered spin anisotropy for spintronic and magnetoelectronic applications. A previous density-functional theory (DFT) study~\cite{PRB-CrCl3} highlighted the importance of including magnetic shape anisotropy, arising from dipole-dipole interactions, to accurately reproduce the in-plane EMA in monolayer CrCl$_3$. However, for CrCl$_3$ and CrBr$_3$, the underlying microscopic mechanisms, particularly the orbital-resolved contributions to the magnetocrystalline anisotropy, remain incompletely understood. Moreover, most studies to date have focused on monolayer systems~\cite{PRB-CrCl3,CrSBr,PRLCrI3,MnX,PRL017201,APLLaBr2,PRBEu,PRBVTe2}, leaving the magnetic anisotropy in the bulk phase relatively unexplored. In particular, a comprehensive analysis of how spin-orbit coupling (SOC), orbital hybridization, and halogen-specific electronic structure collectively influence the EMA in bulk CrCl$_3$ and CrBr$_3$ is still lacking. Addressing these questions may provide valuable insights toward a deeper understanding of magnetic anisotropy in layered vdW magnetic materials.

In this work, we perform first-principles DFT calculations to elucidate the microscopic origin of the EMA in bulk CrCl$_3$ and CrBr$_3$. By systematically analyzing the interplay between the SOC-induced magnetocrystalline anisotropy energy (SOC-MAE) and the shape magnetic anisotropy energy (shape-MAE) arising from magnetic dipole–dipole interactions, we show that the contrasting magnetic behaviors primarily originate from differences in the halogen $p$ orbital characteristics. In CrBr$_3$, greater spatial delocalization of Br 4$p$ orbitals, stronger atomic SOC, and enhanced $p$-$d$ hybridization favor spin-conserving SOC interactions between occupied and unoccupied spin-up $p$ states, particularly yielding a strong positive contribution from the $(p_x, p_y)$ channel. In contrast, the $(p_y, p_z)$ channel contributes negatively but with much smaller magnitude. As a result, the net SOC-MAE remains significantly positive. These effects, amplified by the anisotropic orbital environment, give rise to a sizable out-of-plane SOC-MAE that exceeds the in-plane shape-MAE, thereby stabilizing perpendicular magnetization. In contrast, CrCl$_3$ exhibits more localized Cl 3$p$ orbitals and weaker SOC, which promote spin-flip SOC interactions between opposite-spin states involving both the $(p_x, p_y)$ and $(p_y, p_z)$ channels. These contributions have opposite signs and comparable magnitudes, leading to substantial cancellation due to the more isotropic orbital character. The resulting small SOC-MAE is surpassed by the shape-MAE, favoring in-plane magnetization. These findings highlight the critical role of ligand orbital localization, $p$–$d$ hybridization, and spin-selective SOC interactions in determining magnetic anisotropy, offering microscopic insight for tailoring spin anisotropy in 2D vdW magnets.

%\vspace{0.4cm}
\section{II. CALCULATIONAL METHODS}
%\vspace{0.4cm}

Our first-principles DFT calculations were carried out using the Vienna \textit{ab initio} simulation package (VASP)~\cite{vasp1,vasp2} with a plane-wave basis set. The interactions between core and valence electrons were described using the projector augmented-wave (PAW) method~\cite{paw}. Exchange-correlation effects were treated within the generalized gradient approximation using the Perdew-Burke-Ernzerhof functional~\cite{pbe}. A kinetic energy cutoff of 500~eV and a total energy convergence criterion of $10^{-8}$~eV were employed. Structural relaxations were performed until the Hellmann–Feynman forces on each atom were less than 0.001~eV/\AA. For Brillouin-zone sampling, the $12\times12\times12$ $k$-point mesh was used for both bulk CrCl$_3$ and CrBr$_3$. To account for vdW interactions, the DFT-D3 scheme~\cite{JCP154104,JCC1456} was employed. To properly describe the localized Cr 3$d$ electrons, we used the DFT+$U$ method in the Dudarev formalism~\cite{Dudarev}, with an effective Hubbard $U$ parameter of 3~eV, as suggested in previous studies~\cite{CrSBrguo nanoscale}.

%\vspace{0.4cm}
\section{III. RESULTS and DISCUSSION}
%\vspace{0.4cm}

We begin by optimizing the crystal structures of bulk CrX$_3$ (X = Cl, Br). As shown in Figs.~1(a) and 1(b), both compounds adopt a rhombohedral structure with space group $R\overline{3}$, where each layer consists of edge-sharing CrX$_6$ octahedra forming a honeycomb network of Cr atoms [see Fig.~1(c)]. The optimized lattice parameters are $a = b = 6.003$~\AA{}, $c = 17.311$~\AA{} for CrCl$_3$, and $a = b = 6.360$~\AA{}, $c = 18.341$~\AA{} for CrBr$_3$, with corresponding interlayer distances of $d_{\text{int}} = 3.078$~\AA{} and 3.213~\AA{}, respectively. These values are in good agreement with previous theoretical~\cite{jmcc2015} and experimental studies~\cite{CrCl3jcp,CrBr3jacs} (see Table~I). The larger lattice parameters and interlayer spacing in CrBr$_3$ reflect the larger atomic radius of Br and the relatively weaker Cr--Br bond. Furthermore, Br has a lower electronegativity than Cl, resulting in reduced charge transfer from Cr to Br and a more covalent Cr--Br bond. In contrast, the greater electronegativity difference between Cr and Cl in CrCl$_3$ leads to stronger charge transfer, enhancing the ionic character of the Cr--Cl bond. This increased ionicity tends to localize the Cl $p$ orbitals and weaken $p$--$d$ hybridization with Cr $d$ states. Conversely, in CrBr$_3$, the weaker charge transfer and more delocalized Br 4$p$ orbitals support stronger $p$--$d$ hybridization, which contributes to a narrower band gap, as discussed below.

\begin{figure}
    \centering
    \includegraphics[width=1\linewidth]{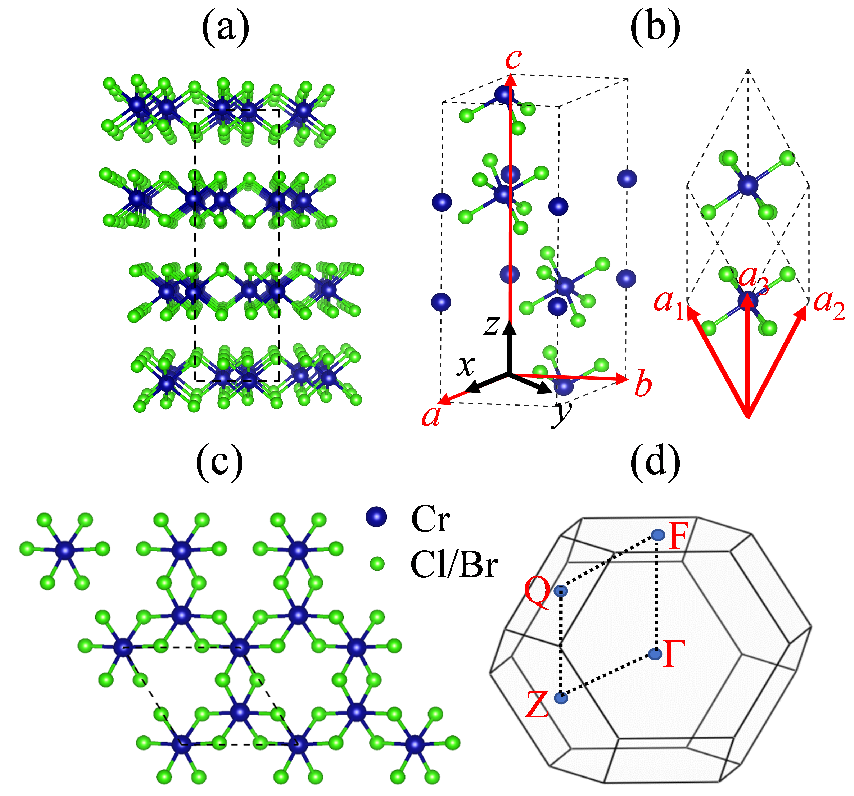}
    \caption{Optimized structure of  bulk CrX$_3$ (X= Cl,Br): (a) perspective view of the layered structure and (b) the conventional unit cell (left) and rhombohedral primitive unit cell (right). The lattice parameters $a$, $b$, and $c$ refer to the conventional cell, whereas $a_1$, $a_2$, and $a_3$ define the rhombohedral primitive cell. The top view of a single CrX$_3$ layer is shown in (c). Blue and green balls represent Cr and Cl (or Br) atoms, respectively. The Brillouin zone corresponding to the rhombohedral primitive unit cell is displayed in (d).}
    \label{fig:enter-label}
\end{figure}

\begin{figure}
    \centering
    \includegraphics[width=1\linewidth]{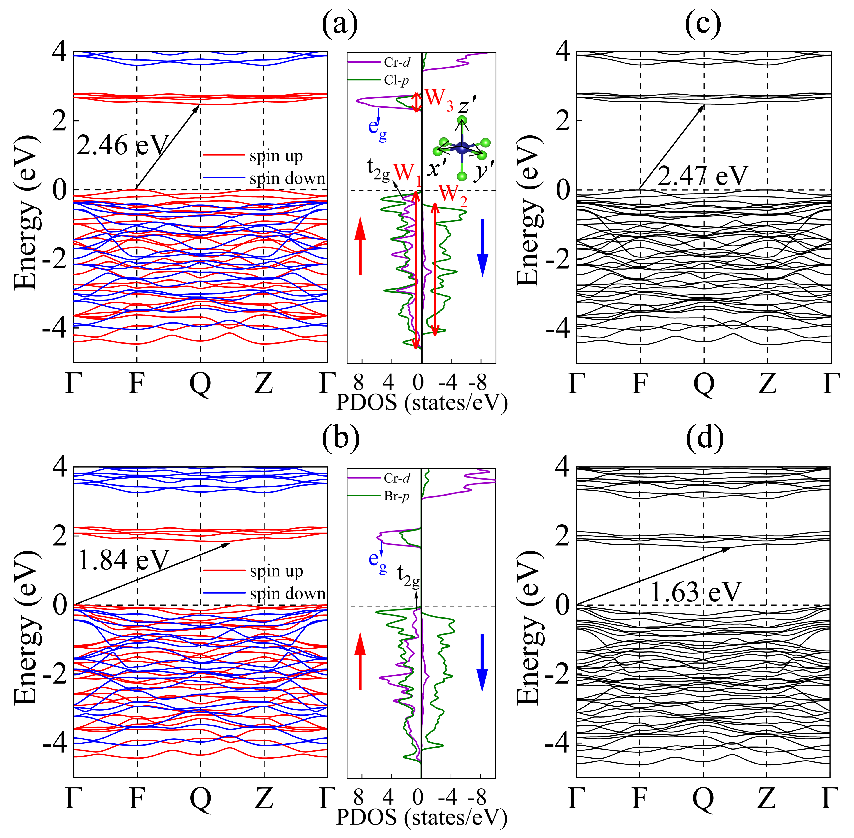}
    \caption{Calculated spin-polarized band structures of (a) CrCl$_3$ and (b) CrBr$_3$ along the high-symmetry path shown in Fig.~1(d), accompanied by the PDOS for Cr $d$ and halogen $p$ orbitals. The Cr $t_{2g}$ and $e_g$ states are defined based on projections onto a local coordinate frame in which the $z'$-axis is aligned with the CrX$_6$ (X = Cl, Br) octahedra, as illustrated in the inset of panel (a). Panels (c) and (d) show the corresponding band structures including SOC for CrCl$_3$ and CrBr$_3$, respectively.}
    \label{fig:enter-label}
\end{figure}

\begin{table}[ht]
\caption{Calculated lattice parameters $a$, $b$, and $c$ for CrCl$_3$ and CrBr$_3$, compared with previous theoretical and experimental results.}
\begin{ruledtabular}
%\scriptsize
\begin{tabular}{llccc}
    &   &  $a$ ({\AA}) & $b$ ({\AA}) & $c$ ({\AA})  \\ \hline
CrCl$_3$  & Present & 6.003 & 6.003  & 17.311 \\
          & Previous~\cite{jmcc2015} & 5.987 & 5.987  & 17.153 \\
          & Experiment~\cite{CrCl3jcp} & 5.942 & 5.942  & 17.333 \\
CrBr$_3$  & Present  & 6.360 & 6.360 & 18.341 \\
          & Previous~\cite{jmcc2015} & 6.350 & 6.350  & 18.259 \\
          & Experiment~\cite{CrBr3jacs} & 6.260 & 6.260  & 18.200 \\

\end{tabular}
\end{ruledtabular}
%\end{ruledtabular}
\end{table}

Figures~2(a) and 2(b) show the spin-polarized band structures and partial density of states (PDOS) of the ferromagnetic phase of CrCl$_3$ and CrBr$_3$, respectively, calculated without SOC. In both compounds, the occupied Cr 3$d$ states are confined almost entirely to the spin-up channel, while the spin-down channel remains largely empty, indicating strong spin polarization. This polarization arises from Hund’s exchange interaction, which favors parallel spin alignment among the three Cr 3$d$ electrons. Combined with the octahedral crystal field from the surrounding halogen ligands [see the inset of Fig.~2(a)], which splits the Cr 3$d$ orbitals into a lower-energy $t_{2g}$ triplet and a higher-energy $e_g$ doublet, this leads to a high-spin $t_{2g}^3e_g^0$ configuration. The spin-up valence bands mainly consist of Cr $t_{2g}$ states hybridized with halogen $p$ orbitals, while the spin-down valence bands are dominated by halogen $p$ character. As summarized in Table~II, the spin-up valence bandwidth $W_1$ is nearly identical in CrCl$_3$ and CrBr$_3$, indicating comparable $t_{2g}$--$p$ hybridization strength. In contrast, the spin-down valence bandwidth $W_2$ is much larger in CrBr$_3$, reflecting the more delocalized nature of Br 4$p$ orbitals compared to Cl 3$p$. The conduction band minimum (CBM) in both materials originates from unoccupied spin-up Cr $e_g$ states, also hybridized with halogen $p$ orbitals. The CBM bandwidth $W_3$ is broader in CrBr$_3$, indicating enhanced $p$--$d$ hybridization driven by the spatially extended Br 4$p$ orbitals.

\begin{table}[ht]
\caption{Calculated bandwidths $W_1$, $W_2$, and $W_3$ [see Fig. 2(a)] corresponding to the spin-up valence band, spin-down valence band, and the CBM, respectively.}
\begin{ruledtabular}
%\scriptsize
\begin{tabular}{lccc}
    &    $W_1$ (eV) &  $W_2$ (eV) &  $W_3$ (eV)   \\ \hline
CrCl$_3$  & 4.74  & 3.94  & 0.65 \\
CrBr$_3$  & 4.71  & 4.17  & 0.75 \\
\end{tabular}
\end{ruledtabular}
%\end{ruledtabular}
\end{table}

Despite their overall qualitative similarity, CrCl$_3$ and CrBr$_3$ exhibit notable differences in their electronic structures, arising from distinct halogen orbital characteristics. In CrCl$_3$, the more localized nature of Cl 3$p$ orbitals limits their spatial overlap with Cr 3$d$ orbitals, resulting in relatively weak $p$-$d$ hybridization. In addition, the higher electronegativity of Cl facilitates greater charge transfer from Cr to Cl, increasing the ionic character of the Cr--Cl bond. This enhanced ionicity leads to stronger electrostatic interaction between the filled Cl $p$ orbitals and the unoccupied Cr $e_g$ orbitals, effectively raising the $e_g$ energy levels relative to the $t_{2g}$ states. As a result, CrCl$_3$ exhibits a larger crystal field splitting, which contributes to its wider band gap. In contrast, CrBr$_3$ features more delocalized Br 4$p$ orbitals and lower electronegativity, which reduce the degree of charge transfer and promote a more covalent Cr--Br bond. The spatially extended Br orbitals enhance $p$--$d$ hybridization with Cr $d$ states, particularly the $e_g$ orbitals, thereby lowering the energy of the CBM. Consequently, CrBr$_3$ experiences a smaller crystal field splitting and a reduced band gap. Our PBE+$U$ calculations yield band gaps of 2.46~eV [see Fig. 2(a)] for CrCl$_3$ and 1.84~eV [Fig. 2(b)] for CrBr$_3$, in good agreement with experimental values of 2.63~eV~\cite{CrCl3nanoscale} and 1.68--2.1~eV, respectively~\cite{CrBr3pccp}. These findings underscore the crucial role of halogen $p$ orbital localization, electronegativity-driven charge transfer, and $p$--$d$ hybridization in modulating the crystal field environment and tuning the electronic properties of layered chromium trihalides.

Furthermore, Figs.~2(c) and 2(d) demonstrate that the inclusion of SOC induces markedly different effects on the band gaps of CrCl$_3$ and CrBr$_3$. In CrCl$_3$, the band gap remains nearly unchanged at 2.47~eV after including SOC [see Fig. 2(c)], indicating a minimal influence. In contrast, CrBr$_3$ exhibits a significant band gap reduction to 1.63~eV [see Fig. 2(d)], corresponding to a decrease of 0.21~eV relative to the SOC-free case. This contrasting behavior originates from the disparity in atomic SOC strength between Cl and Br. Due to its higher atomic number, Br experiences stronger relativistic effects, which lead to appreciable variations in the electronic states near both the valence band maximum and the CBM. The enhanced SOC in Br thus modifies the band edge energies and narrows the gap in CrBr$_3$. By comparison, the lighter Cl atom exhibits much weaker SOC, resulting in only minor perturbations to the band structure of CrCl$_3$ and leaving its band gap essentially unchanged. These findings highlight the importance of halogen atomic SOC in modulating the electronic structure, which ultimately governs the magnetic anisotropy examined below.

\begin{figure}
    \centering
    \includegraphics[width=1\linewidth]{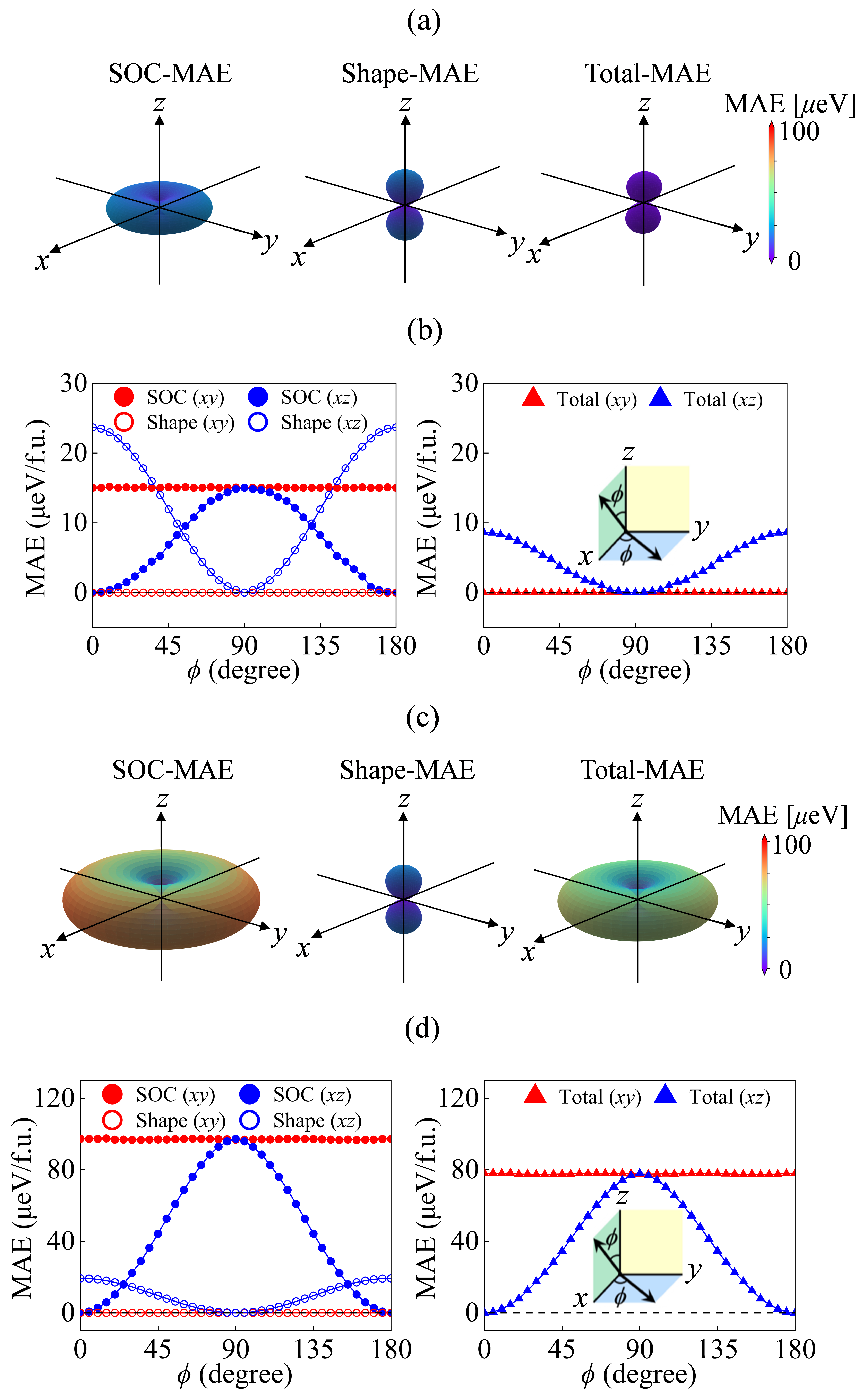}
    \caption{Calculated SOC-MAE, shape-MAE, and total MAE for (a) CrCl$_3$ and (c) CrBr$_3$. The corresponding angular dependence of the energies on the $xy$ and $xz$ planes for CrCl$_3$ and CrBr$_3$ are shown in (b) and (d), respectively, as functions of the angle ${\phi}$ (defined in the inset). The zero-energy reference directions are set to (001) for SOC-MAE, (100) for shape-MAE, and (100) for the total MAE in CrCl$_3$ and (001) for the total MAE in CrBr$_3$.}
    \label{fig:enter-label}
\end{figure}

Next, we investigate the MAE of CrCl$_3$ and CrBr$_3$ by evaluating both the SOC-MAE and the shape-MAE. The SOC-MAE is obtained from the total energy difference between the out-of-plane and in-plane magnetization directions, whereas the shape-MAE arises from long-range dipole-dipole interactions and is given by~\cite{CrSBr,APLLaBr2,PRBEu}:
\begin{equation}\label{eq1}
\small
E^{\rm D} = -\frac{1}{2} \frac{\mu _{0} }{4\pi}\sum_{i\neq j} \frac{1}{r_{ij}^3}\left[\vec{M}_{i}\cdot \vec{M}_{j}-\frac{3}{{r_{ij}^{2}}}\left ( \vec{M}_{i}\cdot \vec{r}_{ij} \right )\left ( \vec{M}_{j}\cdot \vec{r}_{ij} \right )  \right],
\end{equation}
where $\vec{M}_{i}$ and $\vec{M}_{j}$ denote the magnetic moments at atomic sites $i$ and $j$, respectively, $\vec{r}_{ij}$ is the distance vector between the two sites, and $\mu_0$ is the vacuum permeability. The summation is performed over all distinct atomic pairs within a sufficiently large cutoff distance of 1000~{\AA}. Figures~3(a) and 3(c) display the calculated SOC-MAE, shape-MAE, and their sum (total MAE) for CrCl$_3$ and CrBr$_3$, respectively. In CrCl$_3$, the SOC-MAE favors out-of-plane magnetization with a value of $15$~$\mu$eV, while the shape-MAE favors in-plane alignment with $23$~$\mu$eV [see Fig.~3(b)]. Since the shape-MAE slightly exceeds the SOC-MAE, the resulting EMA lies in-plane with a total MAE value of $8$~$\mu$eV. In contrast, CrBr$_3$ exhibits a significantly larger SOC-MAE of $97$~$\mu$eV that favors out-of-plane magnetization, exceeding the shape-MAE of $20$~$\mu$eV that favors in-plane alignment [see Fig.~3(d)]. As a result, CrBr$_3$ exhibits an out-of-plane EMA with a total MAE value of $77$~$\mu$eV. Notably, the SOC-MAE in CrBr$_3$ is more than six times greater than that in CrCl$_3$, which can be attributed to the stronger SOC of Br atoms and the anisotropic nature of Br 4$p$ orbitals, as discussed below.

\begin{figure}
    \centering
    \includegraphics[width=1\linewidth]{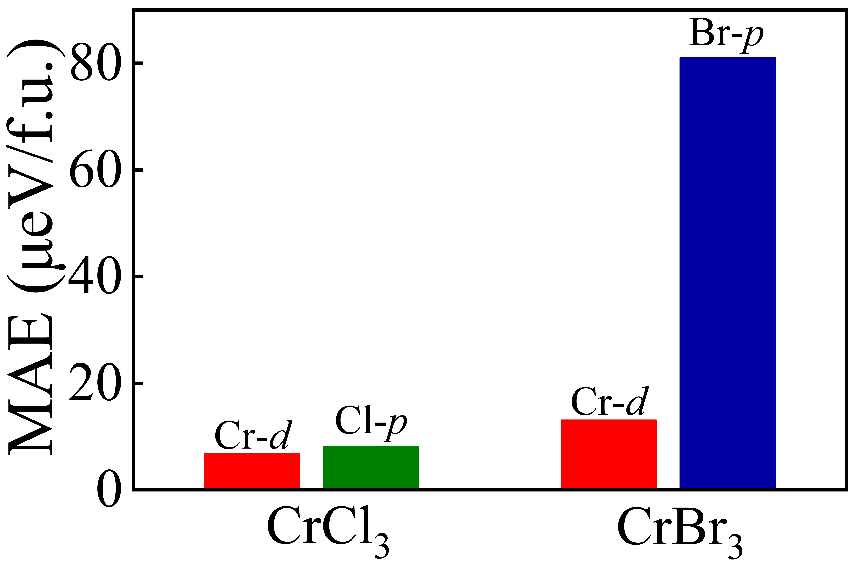}
    \caption{Calculated atom- and orbital-resolved contributions to the SOC-MAE in CrCl$_3$ and CrBr$_3$.}
    \label{fig:enter-label}
\end{figure}

\begin{figure}
    \centering
    \includegraphics[width=1\linewidth]{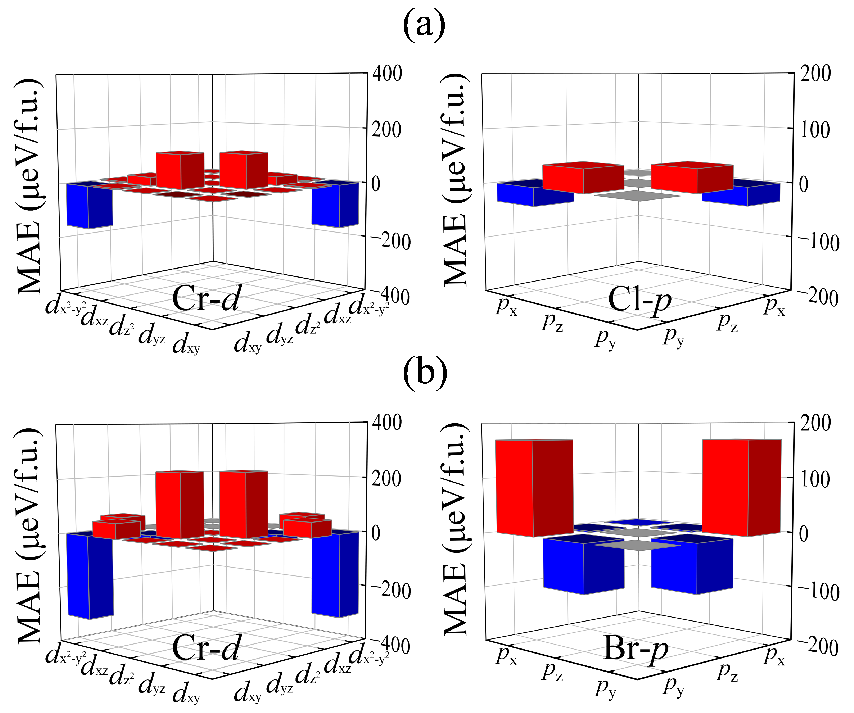}
    \caption{Interorbital coupling contributions to the SOC-MAE in (a) CrCl$_3$ and (b) CrBr$_3$, involving Cr $d$ and halogen $p$ orbitals. Positive (negative) values indicate a preference for out-of-plane (in-plane) magnetic anisotropy.}
    \label{fig:enter-label}
\end{figure}

To elucidate the substantial difference in SOC-MAE between CrCl$_3$ and CrBr$_3$, we analyze the atom- and orbital-resolved contributions, as shown in Fig. 4. Both compounds display comparable contributions from Cr $d$ orbitals, with CrBr$_3$ showing only a slight enhancement. However, this marginal increase alone cannot account for the pronounced disparity in their overall SOC-MAE. The key difference originates from the halogen atoms. In CrCl$_3$, the contribution of Cl $p$ orbitals to the SOC-MAE is similar in magnitude to that of the Cr $d$ orbitals, resulting in a relatively modest total SOC-MAE. In contrast, CrBr$_3$ exhibits a significantly larger SOC-MAE, primarily due to a dominant contribution from Br $p$ orbitals. These orbitals exhibit strong magnetic anisotropy that favors out-of-plane magnetization, thereby playing a crucial role in enhancing the SOC-MAE. While the final orientation of the EMA is governed by the competition between SOC-MAE and shape-MAE, it is the substantial Br $p$ orbital contribution that fundamentally drives the contrasting magnetic anisotropies of CrCl$_3$ and CrBr$_3$.

Figures 5(a) and 5(b) present a detailed breakdown of the orbital-resolved SOC-MAE by illustrating the interorbital coupling contributions in CrCl$_3$ and CrBr$_3$, respectively. In both compounds, the Cr $d$ orbitals exhibit two dominant interorbital contributions: a negative term from the ($d_{xy}$, $d_{x^2 - y^2}$) pair and a positive term from the ($d_{z^2}$, $d_{yz}$) pair. Although these competing contributions largely offset each other, additional minor positive terms from other $d$-orbital channels result in a net positive SOC-MAE, with a slightly larger value in CrBr$_3$ (see Fig.~4). In contrast, the halogen $p$ orbitals reveal a striking difference between the two compounds, both in the sign and magnitude of the key coupling terms. In CrCl$_3$, the dominant contributions arise from the ($p_x$, $p_y$) and ($p_y$, $p_z$) channels. The ($p_x$, $p_y$) interaction contributes negatively, while the ($p_y$, $p_z$) term contributes positively, and their magnitudes are similar. This near-symmetric cancellation leads to a relatively small net SOC-MAE from the Cl $p$ orbitals, comparable to the Cr $d$ contribution (see Fig. 4). In CrBr$_3$, however, the $(p_x,p_y)$ coupling is strongly positive and significantly larger, whereas the $(p_y,p_z)$ contribution is negative and much smaller in magnitude than the $(p_x,p_y)$ term. This inversion in sign and imbalance in strength suppresses the mutual cancellation found in CrCl$_3$, allowing the Br $p$ orbitals to contribute a substantial net positive SOC-MAE. As a result, the Br $p$ states give rise to a pronounced out-of-plane magnetic anisotropy, as evident in Fig.~4.

\begin{figure}
    \centering
    \includegraphics[width=1\linewidth]{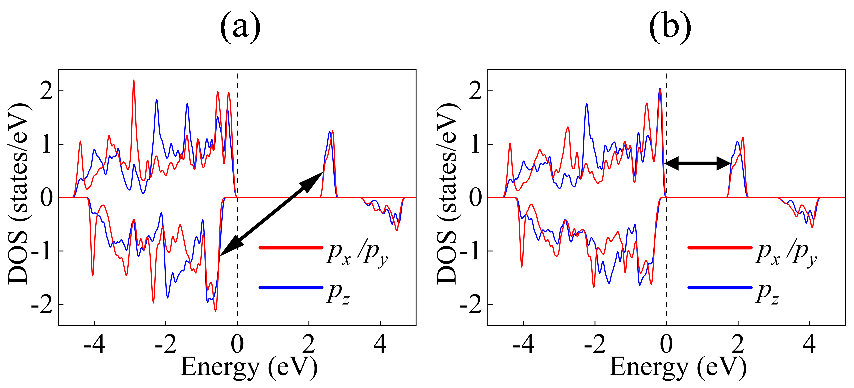}
    \caption{Calculated PDOS for halogen $p$ orbitals in (a) CrCl$_3$ and (b) CrBr$_3$. The arrows in (a) and (b) indicate spin-flip and spin-conserving SOC interactions in CrCl$_3$ and CrBr$_3$, respectively.}
    \label{fig:enter-label}
\end{figure}

To understand the contrasting behavior of halogen $p$ orbitals in determining the SOC-MAE of CrCl$_3$ and CrBr$_3$, we analyze the results within the framework of second-order perturbation theory~\cite{PRBEu}, where the SOC-MAE is expressed as
\begin{equation}\label{eq2}
\small
\begin{split}
{\rm MAE}
 &= \xi^2\sum_{\sigma \sigma^{\prime}} \sum_{o^{\sigma}, u^{\sigma^{\prime}}}
 \frac{
 \left[
 (2\delta_{\sigma \sigma^{\prime}} - 1)
 \left(
 \left| \left\langle o^{\sigma} \left| L_z \right| u^{\sigma^{\prime}} \right\rangle \right|^2
 -
 \left| \left\langle o^{\sigma} \left| L_x \right| u^{\sigma^{\prime}} \right\rangle \right|^2
 \right)
 \right]}
 {\varepsilon_{u}^{\sigma^{\prime}} - \varepsilon_{o}^{\sigma}},
\end{split}
\end{equation}
where $\xi$ is the SOC constant, and $\varepsilon_{o}^{\sigma}$ ($\varepsilon_{u}^{\sigma^{\prime}}$) denote the energies of occupied (unoccupied) states with spin $\sigma$ ($\sigma^{\prime}$). The numerator in Eq.~\eqref{eq2} represents the the difference of SOC matrix elements, as summarized in Table III. As shown in Fig.~5(a), the dominant halogen contributions to the MAE in CrCl$_3$ stem from the $(p_x, p_y)$ and $(p_y, p_z)$ channels, which contribute negatively and positively, respectively. These contributions are primarily driven by spin-flip SOC interactions ($\sigma \ne \sigma'$) between occupied spin-down and unoccupied spin-up states, as illustrated in Fig.~6(a) and detailed in Table~III. Since the two channels contribute with opposite signs and similar magnitudes, their effects largely cancel out, resulting in a reduced net SOC-MAE from the Cl $p$ orbitals. In contrast, Fig.~5(b) shows that in CrBr$_3$, the SOC-MAE is dominated by a strong positive contribution from the $(p_x, p_y)$ channel, while the $(p_y, p_z)$ channel is negative with much smaller magnitude. These terms are mostly driven by spin-conserving interactions ($\sigma = \sigma'$) between spin-up states, as shown in Fig.~6(b) and Table III. The resulting asymmetry reduces cancellation, enhancing the net SOC-MAE associated with the Br $p$ orbitals. This contrasting behavior can be attributed to fundamental differences in orbital character, SOC strength, and $p$--$d$ hybridization. In CrCl$_3$, the localized, isotropic nature of the Cl $3p$ orbitals and relatively weak SOC promote spin-flip processes mediated by the $L^+S^-$ and $L^-S^+$ terms in the SOC Hamiltonian, $H_{\text{SOC}} = \xi \mathbf{L} \cdot \mathbf{S}$. However, the near-equal magnitudes and opposing signs of contributions across different channels lead to destructive interference, suppressing the overall anisotropy. Conversely, CrBr$_3$ features more delocalized Br $4p$ orbitals, a larger SOC constant, and stronger $p$--$d$ hybridization. These factors favor spin-conserving SOC interactions—especially in the $(p_x, p_y)$ channel—which benefit from strong orbital anisotropy and minimal cancellation. Consequently, a pronounced positive SOC-MAE emerges, stabilizing an out-of-plane easy axis. Taken together, these results show that while CrCl$_3$ exhibits reduced SOC-MAE due to competing spin-flip interactions that partially cancel each other, CrBr$_3$ achieves a robust anisotropic response driven by dominant spin-conserving interactions, which reflect the intrinsic electronic character of Br $p$ orbitals and their enhanced coupling to Cr $d$ states.

\begin{table}[ht]
\caption{The SOC matrix elements for halogen $p$ orbitals~\cite{SOC-P}. $o$ and $u$ represent the occupied and unoccupied states, respectively.}
\begin{ruledtabular}
%\scriptsize
\begin{tabular}{lrrr}
    & $u^{\sigma^{\prime} = \uparrow}$ ($p_y$) & $u^{\sigma^{\prime} = \uparrow}$ ($p_z$) & $u^{\sigma^{\prime} = \uparrow}$ ($p_x$)  \\ \hline
$o^{\sigma = \downarrow}$ ($p_y$) & 0 & 1 & $-$1  \\
$o^{\sigma = \downarrow}$ ($p_z$) & 1 & 0 & 0   \\
$o^{\sigma = \downarrow}$ ($p_x$) & $-$1 & 0 &  0  \\ \hline
$o^{\sigma = \uparrow}$ ($p_y$) & 0 & $-$1 & 1    \\
$o^{\sigma = \uparrow}$ ($p_z$) & $-$1 & 0  & 0 \\
$o^{\sigma = \uparrow}$ ($p_x$) & 1 & 0 & 0 \\
 \end{tabular}
\end{ruledtabular}
%\end{ruledtabular}
\end{table}

%\vspace{0.4cm}
\section{IV. SUMMARY}
%\vspace{0.4cm}

Our first-principles calculations elucidate the microscopic origin of the contrasting magnetic anisotropy in bulk CrCl$_3$ and CrBr$_3$. While both compounds exhibit comparable out-of-plane contributions from Cr $d$ orbitals, the key difference arises from the halogen $p$ orbitals. In CrCl$_3$, the weak SOC and localized Cl 3$p$ orbitals facilitate spin-flip interactions that yield contributions of opposite sign across multiple $p$--$p$ channels. These competing terms, similar in magnitude but opposite in sign, lead to substantial cancellation and thus a greatly reduced net SOC-MAE, consistent with an isotropic orbital angular momentum environment. As a result, the shape-MAE dominates, stabilizing an in-plane EMA. In contrast, CrBr$_3$ exhibits stronger SOC and more delocalized Br 4$p$ orbitals, which promote spin-conserving interactions—particularly in the $(p_x, p_y)$ channel—within a highly anisotropic orbital environment. These interactions produce a large, positive SOC-MAE that surpasses the shape-MAE, favoring an out-of-plane EMA. These findings underscore the crucial role of ligand SOC strength, orbital anisotropy, and spin selection rules in determining magnetic anisotropy, providing a pathway for engineering spin–orbit-driven functionalities in 2D vdW magnetic semiconductors.

\vspace{0.4cm}

\noindent {\bf Acknowledgements.}
This work was supported by the the Natural Science Foundation of Henan (No. 252300421216), the talent Introduction Project in Henan Province (HNGD2025008), the Innovation Program for Quantum Science and Technology (No. 2021ZD0302800), the International Cooperation Project of Science and Technology of Henan Province (No. 242102520029), and the Foundation of Henan Educational Committee (No. 25A140003). J.H.C acknowledges the support from the National Research Foundation of Korea (NRF) grant funded by the Korean Government (Grant No. RS202300218998). The calculations were performed by the KISTI Supercomputing Center through the Strategic Support Program (Program No. KSC-2024-CRE-0055) for the supercomputing application research.

                  %%%%%  REFERENCES  %%%%%

\noindent Z. S. and S. L. contributed equally to this work \\
\noindent $^{*}$ Corresponding authors: wb@henu.edu.cn and cho@henu.edu.cn


\begin{thebibliography}{99}

\bibitem{NatureCrI3} B. Huang, G. Clark, E. Navarro-Moratalla, D. R. Klein, R. Cheng, K. L. Seyler, D. Zhong, E. Schmidgall, M. A. McGuire, D. H. Cobden, W. Yao, D. Xiao, P. Jarillo-Herrero and X. Xu, Layer-dependent ferromagnetism in a van der Waals crystal down to the monolayer limit, Nature, 546 (2017) 270-273.
\bibitem{NatureCrGeTe3} C. Gong, L. Li, Z. Li, H. Ji, A. Stern, Y. Xia, T. Cao, W. Bao, C. Wang, Y. Wang, Z. Q. Qiu, R. J. Cava, S. G. Louie, J. Xia and X. Zhang, Discovery of intrinsic ferromagnetism in two-dimensional van der Waals crystals, Nature, 546 (2017) 265-269.
\bibitem{3} H. Kurebayashi, J. H. Garcia, S. Khan, J. Sinova and S. Roche, Magnetism, symmetry and spin transport in van der Waals layered systems, Nat. Rev. Phys., 4 (2022) 150-166.
\bibitem{natureFeGeTe}Y. Deng, Y. Yu, Y. Song, J. Zhang, N. Z. Wang, Z. Sun, Y. Yi, Y. Z. Wu, S. Wu and J. Zhu, Gate-tunable room-temperature ferromagnetism in two-dimensional Fe$_3$GeTe$_2$, Nature, 563 (2018) 94-99.
\bibitem{2d vdw} K. S. Burch, D. Mandrus and J.-G. Park, Magnetism in two-dimensional van der Waals materials, Nature, 563 (2018) 47-52.
\bibitem{vse2} M. Bonilla, S. Kolekar, Y. Ma, H. C. Diaz, V. Kalappattil, R. Das, T. Eggers, H. R. Gutierrez, M.-H. Phan and M. Batzill, Strong room-temperature ferromagnetism in VSe$_2$ monolayers on van der Waals substrates, Nat. Nanotech., 13 (2018) 289-293.
\bibitem{NC FeGeTe} Z. Fei, B. Huang, P. Malinowski, W. Wang, T. Song, J. Sanchez, W. Yao, D. Xiao, X. Zhu and A. F. May, Two-dimensional itinerant ferromagnetism in atomically thin Fe$_3$GeTe$_2$, Nat. Mater. 17 (2018) 778-782.
\bibitem{NM 2d} S. Jiang, J. Shan and K. F. Mak, Electric-field switching of two-dimensional van der Waals magnets, Nat. Mater., 17 (2018) 406-410.
\bibitem{2d magnetic} X. Jiang, Q. Liu, J. Xing, N. Liu, Y. Guo, Z. Liu and J. Zhao, Recent progress on 2D magnets: Fundamental mechanism, structural design and modification, Appl. Phys. Rev., 8 (2021) 031305.
\bibitem{Nanotechnol-2019} M. Gibertini, M. Gibertini, A. F. Morpurgo  and K. S. Novoselov, Magnetic 2D materials and heterostructures, Nat. Nanotech., 14 (2019) 408-419.
\bibitem{science-2019} C. Gong and X. Zhang, Two-dimensional magnetic crystals and emergent heterostructure devices, Science, 363 (2019) eaav4450.
\bibitem{PRL-MW1966} N. D. Mermin, H. Wagner, Absence of ferromagnetism or antiferromagnetism in one- or two-dimensional isotropic heisenberg models, Phys. Rev. Lett., 17 (1966) 1133-1136.
\bibitem{AMCrCl3-xBrx} M. Abramchuk, S. Jaszewski, K. R. Metz, G. B. Osterhoudt, Y. Wang, K. S. Burch and F. Tafti, Controlling Magnetic and Optical Properties of the van der Waals Crystal CrCl$_{3-x}$Br$_x$ via Mixed Halide Chemistry, Adv. Mater., 30 (2018) 1801325.
\bibitem{NL-CrCl3}   X. Cai, T. Song, N.P. Wilson, G. Clark, M. He, X. Zhang, T. Taniguchi, K. Watanabe, W. Yao, D. Xiao, M.A. McGuire, D.H. Cobden and X. Xu, Atomically Thin CrCl$_3$: An In-Plane Layered Antiferromagnetic Insulator, Nano. Lett., 19 (2019) 3993-3998
\bibitem{PNAS-CrX3} H.H. Kim, B. Yang, S. Li, S. Jiang, C. Jin, Z. Tao, G. Nichols, F. Sfigakis, S. Zhong, C. Li, S. Tian, D.G. Cory, G.X. Miao, J. Shan, K.F. Mak, H. Lei, K. Sun, L. Zhao and A.W. Tsen, Evolution of interlayer and intralayer magnetism in three atomically thin chromium trihalides, Proc. Natl. Acad. Sci. U S A, 116 (2019) 11131-11136.
\bibitem{PRB-CrCl3} F. Xue, Y. Hou, Z. Wang, and R. Wu, Two-dimensional ferromagnetic van der Waals CrCl$_3$ monolayer with enhanced anisotropy and Curie temperature, Phys. Rev. B, 100 (2019) 224429.
 \bibitem{CrSBr} B. Wang, Y. Wu, Y. Bai, P. Shi, G. Zhang, Y. Zhang, and C. Liu, Origin and regulation of triaxial magnetic anisotropy in the ferromagnetic semiconductor CrSBr monolayer, Nanoscale, 15 (2023) 13402.
\bibitem{PRLCrI3} D.H. Kim, K. Kim, K.T. Ko, J. Seo, J.S. Kim, T.H. Jang, Y. Kim, J.Y. Kim, S.W. Cheong and J.H. Park, Giant Magnetic Anisotropy Induced by Ligand LS Coupling in Layered Cr Compounds, Phys. Rev. Lett., 122 (2019) 207201.
\bibitem{MnX} B. Wang, Y. Zhang, L. Ma, Q. Wu, Y. Guo, X. Zhang and J. Wang, MnX (X = P, As) monolayers: a new type of two-dimensional intrinsic room temperature ferromagnetic half-metallic material with large magnetic anisotropy, Nanoscale, 11 (2019) 4204-4209.
\bibitem{PRL017201} I. Lee, F.G. Utermohlen, D. Weber, K. Hwang, C. Zhang, J. van Tol, J.E. Goldberger, N. Trivedi and P.C. Hammel, Fundamental Spin Interactions Underlying the Magnetic Anisotropy in the Kitaev Ferromagnet CrI$_3$, Phys. Rev. Lett., 124 (2020) 017201.
\bibitem{APLLaBr2} J. Zhang, Y. Shao, C. Li, J. Xu, H. Zhang, C. Wang, B. Wang and J. Cho, Nonvolatile electrical control of magnetic anisotropy in ferromagnetic LaBr$_2$ monolayer on ferroelectric In$_2$Se$_3$ substrate, Appl. Phys. Lett., 125 (2024) 142404.
\bibitem{PRBEu} B. Wang, Y. Bai, C. Wang, S. Liu, S. Yao, Y. Jia and J. Cho, Ferroelectric control of magnetic anisotropy in multiferroic heterostructure EuSn$_2$As$_2$/In$_2$Se$_3$, Phys. Rev. B, 110 (2024) 094423.
\bibitem{PRBVTe2} Z. Tang, Y. Chen, Y. Zheng and X. Luo, Strain engineering magnetocrystalline anisotropy in strongly correlated VTe$_2$ with room-temperature ferromagnetism, Phys. Rev. B, 105 (2022) 214403.

\bibitem{vasp1} G. Kresse and J. Hafner, Ab initio molecular dynamics for open-shell transition metals, Phys. Rev. B,  48 (1993) 13115.
\bibitem{vasp2} G. Kresse and J. Furthm\"uller, Efficiency of ab-initio total energy calculations for metals and semiconductors using a plane-wave basis set, Comput. Mater. Sci.,  6 (1996) 15.
\bibitem{paw} P. E. Bl\"ochl, Projector augmented-wave method, Phys. Rev. B 50 (1994) 17953.
\bibitem{pbe} J. P. Perdew, K. Burke, and M. Ernzerhof, Generalized gradient approximation made simple, Phys. Rev. Lett.,  77 (1996) 3865; 78 (1997) 1396(E).
\bibitem{JCP154104} S. Grimme, J. Antony, S. Ehrlich and H. Krieg, A Consistent and Accurate ab initio Parametrization of Density Functional Dispersion Correction (DFT-D) for the 94 Elements H-Pu, J. Chem. Phys., 132 (2010) 154104.
\bibitem{JCC1456} S. Grimme, S. Ehrlich and L. Goerigk, Effect of the Damping Function in Dispersion Corrected Density Functional Theory, J. Comput. Chem.,32 (2011) 1456.
\bibitem{Dudarev} S. L. Dudarev, G. A. Botton, S. Y. Savrasov, C. J. Humphreys and A. P. Sutton, Electron-energy-loss spectra and the structural stability of nickel oxide: An LSDA+U study, Phys. Rev. B, 57 (1998) 1505.
\bibitem{CrSBrguo nanoscale} Y. Guo, Y. Zhang, S. Yuan, B. Wang and J. Wang, Chromium sulfide halide monolayers: intrinsic ferromagnetic semiconductors with large spin polarization and high carrier mobility, Nanoscale, 10 (2018) 18036-18042.
\bibitem{jmcc2015} W. Zhang, Q. Qu, P. Zhu and C.-H. Lam, Robust intrinsic ferromagnetism and half semiconductivity in stable two-dimensional single-layer chromium trihalides, J. Mater. Chem. C, 3 (2015) 12457-12468.
\bibitem{CrCl3jcp} B. Morosin and A. Narath, X‐ray diffraction and nuclear quadrupole resonance studies of chromium trichloride, J. Chem. Phys., 40 (1964) 1958-1967.
\bibitem{CrBr3jacs} L. Handy and N. Gregory, Structural properties of chromium (III) iodide and some chromium (III) mixed halides, J. Am. Chem. Soc., 74 (1952) 891-893
\bibitem{CrCl3nanoscale} D. Mastrippolito, L. Ottaviano, J. Wang, J. Yang, F. Gao, M. Ali, G. D'Olimpio, A. Politano, S. Palleschi and S. Kazim, Emerging oxidized and defective phases in low-dimensional CrCl$_3$, Nanoscale. Adv., 3 (2021) 4756-4766.
\bibitem{CrBr3pccp} D. Baral, Z. Fu, A. S. Zadorozhnyi, R. Dulal, A. Wang, N. Shrestha, U. Erugu, J. Tang, Y. Dahnovsky and J. Tian, Small energy gap revealed in CrBr$_3$ by scanning tunneling spectroscopy, Phys. Chem. Chem. Phys., 23 (2021) 3225-3232.


\bibitem{SOC-P}  Q. Cui, Y. Zhu, J. Jiang, J. Liang, D. Yu, P. Cui, and H. Yang, Ferroelectrically controlled topological magnetic phase in a Janus-magnet-based multiferroic heterostructure, Phys. Rev. Res., 3 (2021) 043011.



\end{thebibliography}
\end{document}